\documentclass[reprint,onecolumn,amsmath,amssymb,superscriptaddress,11pt]{revtex4-1}
\usepackage{graphicx}
\usepackage{bm}
\usepackage[utf8]{inputenc}
\usepackage{paralist}
\usepackage{graphics}
\usepackage{epsfig}

\begin{document}

\relpenalty=9999
\binoppenalty=9999
\newcommand{\N} {\mathbb N}
\newcommand{\abs}[1]{\left|{#1}\right|}
\def \ind{1{\hskip -2.5 pt}\hbox{I}}
\newcommand{\ER} {Erd\H{o}s-R\'enyi }
\newcommand{\prob}[1]{\ensuremath{\mathbf{P}\big(\,#1\,\big)}}
\newcommand{\expect}[1]{\ensuremath{\mathbf{E}\big(\,#1\,\big)}}
\newcommand{\expt}[1]{\ensuremath{\mathbf{E}(#1)}}
\newcommand{\var}[1]{\ensuremath{\mathbf{D}^2\left(#1\right)}}
\newcommand{\cov}[2]{\ensuremath{\mathbf{Cov}\big(\,#1\,,\,#2\,\big)}}
\newcommand{\condprob}[2]{\ensuremath{\mathbf{P}\big(\,#1\,\big|\,#2\,\big)}}
\newcommand{\condexpect}[2]{\ensuremath{\mathbf{E}\big(\,#1\,\big|\,#2\,\big)}}
\newcommand{\condvar}[2]{\ensuremath{\mathbf{D}^2\left(#1\,\big|\,#2\right)}}
\newcommand{\condcov}[3]{\ensuremath{\mathbf{Cov}\big(\,#1\,,\,#2,\,\big|\,#3\,\big)}}

\title{An equation-free approach \\
to coarse-graining the dynamics of networks}

\author{Katherine A. Bold}
\email{kbold@math.princeton.edu}
\affiliation{Program in Applied and Computational Mathematics (PACM), Princeton University, Princeton, New Jersey 08544, USA\footnote{Present address: CUSP, University of Washington at Bothell, Bothell WA, USA.  }}
\author{Karthikeyan Rajendran}
\email{krajendr@princeton.edu}
\affiliation{Department of Chemical and Biological Engineering, Princeton University, Princeton, New Jersey 08544, USA}
\author{Bal\'azs R\'ath}
\email{rathb@math.bme.hu}
\affiliation{Institute of Mathematics, Budapest University of Technology (BME), H-1111 Budapest, Hungary.\footnote{Present address: Dept. of Mathematics, ETH Zurich, 8092 Zurich, Switzerland.}}
\author{Ioannis G. Kevrekidis}
\email{yannis@princeton.edu}
\affiliation{Department of Chemical and Biological Engineering, Princeton University, Princeton, New Jersey 08544, USA}
\affiliation{Program in Applied and Computational Mathematics (PACM), Princeton University, Princeton, New Jersey 08544, USA}

\date{\today}

\begin{abstract}
We propose and illustrate an approach to coarse-graining the dynamics of evolving networks
(networks whose connectivity changes dynamically).
The approach is based on the equation-free framework: short bursts of
detailed network evolution simulations are coupled with lifting and
restriction operators that translate between actual network realizations
and their (appropriately chosen) coarse observables.
This framework is used here to accelerate temporal simulations (through
coarse projective integration),
and to implement coarse-grained fixed point algorithms (through matrix-free Newton-Krylov GMRES).
The approach is illustrated through a simple network evolution example,
for which analytical approximations to the coarse-grained dynamics can
be independently obtained, so as to validate the computational results.
The scope and applicability of the approach, as well as the issue of
selection of good coarse observables are discussed.
\end{abstract}

\maketitle

\section{\label{sec:intro}Introduction}

Complex dynamic systems, exhibiting emergent dynamics, often arise
in the form of graphs (or networks):  the internet, social networks,
chemical and biochemical reaction networks, communication networks and more
\cite{Falo99power-law,Albe02statistical,Newm02random,Newm03structure,Aren06synchronizationa,
Bocc06complex,Newm06structure,Barr08dynamical,Binz09topology,Lain09dynamics,Toiv09comparative}.
In a social network, for example, the individuals are represented by nodes (or vertices),
while the relations among them are represented by the edges connecting these nodes.

One type of network dynamics arises in cases where
the network topology (connectivity) is static, but the {\em state} of each
vertex is a variable that evolves in time (in part due to interactions with
the states of connected vertices).
Such problems are said to exhibit ``dynamics {\em on} networks".
A different type of network dynamics (on which we focus here) arises when
the {\em existence} or the {\em strength} of connections between the
vertices constitute the variables that evolve in time.
These problems are said to exhibit ``dynamics {\em of} networks".
These two types of dynamics are not, of course, mutually exclusive; clearly
we can have dynamical problems involving dynamics both {\em on} and {\em of} networks.
The evolutionary network problems we will refer to in this work will be
exclusively of the second type of dynamics mentioned above: dynamics {\em of} networks,
where the network structure is the state that changes over time.
We will restrict ourselves to networks with unweighted edges, that are either
present or absent; we will not study edges of continuously variable strength
here, even though our methods can be adapted to function in such situations also
(in fact, with appropriate modifications, in any type of network evolution problem).

In our networks of choice the detailed graph representation
constantly changes over time according to some specified rules:
edges (and/or nodes) are added or deleted.
Although these fine-scale, ``node level",  microscopic details of the graph are changing in time,
coherence often emerges at a macroscopic level.
At such a coarse-grained, system scale, the (expected) structure is often seen to evolve smoothly in time:
it may eventually become stationary or may possibly oscillate between a number of states.
We will use a coarse-grained, macroscopic view to study graph dynamics,
treating the coarse temporal evolution as a dynamical system.

Reduced descriptions of high-dimensional dynamical systems (in our case, of large graphs)
are possible predominantly due
to two mechanisms: (a) {\em decoupling} of the evolution of a set of variables and/or (b)
{\em separation of time scales} between evolution of different groups of variables.
In the former (much simpler) case, the few uncoupled variables evolve {\em independently} of other
variations in the system and hence a reduced closed description can be written
in terms of just these variables.
In the latter (more interesting) case, characteristic time scales of evolution of a few variables
(called the ``slow" variables) of the system are much longer than the time scales
of evolution of the remaining ``fast" variables.
After a short evolution time the fast variables will then often become {\em slaved to the slow variables}, i.e.,
the evolution of the fast variables becomes (in the long term) solely determined by the evolution of the
slow variables.
The long-term dynamics of the system can therefore in principle be approximated by equations
written only in terms of the slow variables, the ``coarse variables" of the system.
Note also that, depending on the time scales of interest, it may be possible to close the system
(write closed form evolution equations) at different levels of detail.

%
To use the established tools and techniques of dynamical systems (e.g.
bifurcation and stability computations), however, one must
explicitly derive the evolution equations describing the coarse
variable dynamics.
We are interested in cases where such coarse evolution equations {\em conceptually exist},
yet they are not available in closed form.
In such cases, it has been shown that it may be possible to {\em circumvent}
their explicit derivation using equation-free techniques \cite{Kevr03equation-free,Kevr04equation-free:}.
These techniques are based on short bursts of appropriately designed computational experiments with the detailed,
``fine scale" network dynamic evolution model, and on the knowledge of the appropriate {\em coarse observables}:
the variables in terms of which reduced closed coarse equations could {\em in principle} be written.
Using traditional numerical algorithms as the basis for the design of computational experiments,
and exploiting algorithms that translate between coarse variable values (relevant network statistics)
and actual realizations of networks with these statistics, equation-free approaches may
significantly accelerate the computer-assisted study of network dynamics.
In recent work, we have applied equation-free techniques
to illustrative examples of several different types: molecular dynamics \cite{Chen04from},
collective animal behavior \cite{Moon07heterogeneous},
cell population dynamics \cite{Bold07equation-free} and
also dynamical models on static networks \cite{Raje11coarse}.
Here, we demonstrate the use of equation-free techniques on an illustrative
graph evolution problem, and test our results against explicitly
derived coarse equations.
%

%

A crucial prerequisite for equation-free modeling is the knowledge of
a good set of {\em coarse variables} - the collective network features that
can be used to predict its (expected) coarse evolution, the variables
in terms of which the coarse model would close.
While a large graph is an
intrinsically high dimensional object and difficult
to visualize, its complicated structure can be probed
by measuring  statistical properties of the graph.
Such statistical properties of graphs are often good
candidate coarse variables.
Commonly used statistical properties for describing
a graph include the average degree, the degree distribution, the clustering coefficient,
and the distributions of shortest path lengths.
\cite{Newm03structure}.

As we will see below, even after a good set of such coarse variables has been
selected, an important requirement for using our equation-free algorithms is the
ability to routinely construct graphs exhibiting prescribed values of these
properties.
It is clearly trivial to construct a graph with a given number
of nodes and edges;  yet it is quite non-trivial to construct realizations
of graphs with, say,  a prescribed distribution of shortest path lengths.
%
%
%

These issues will be discussed and illustrated through a simple model example
in the remainder of the paper as follows:
The model behavior is discussed briefly in Section~\ref{sec:model}.
We describe our coarse-graining approach in Section~\ref{sec:cg}.
For our simple example, it so happens that several theoretical results can be
explicitly obtained (and used for validation of the computer-assisted approach).
These will be discussed in Sections~\ref{sec:theory} and \ref{sec:addres}.
We will conclude with a brief summary and a discussion of the scope of the approach, its
strengths and shortcomings, and of certain (in our opinion) important open research directions.

\section{\label{sec:model}Model: A random evolution of networks}

We consider a simple, illustrative model of stochastic network evolution.
Let the graph at any discrete time, $T = 0, 1,...$ be denoted by $G_n(T)$.
The subscript $n$ denotes the number of nodes in the network.
The rules governing the dynamics at each iteration can be described as follows:

\begin{enumerate}
\item Select a pair of nodes at random and connect them together by an edge if they
are not already connected to each other.
\item With probability $r$, remove an edge chosen uniformly at random.
($r$ stands for {\em removal probability}.)
\end{enumerate}

\begin{figure}
\includegraphics[width=0.5\textwidth]{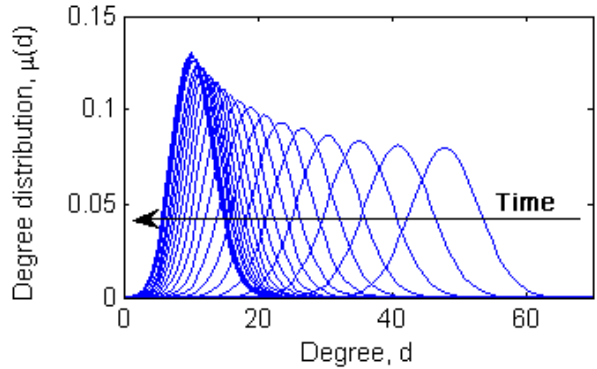}
\caption{\label{fig:dd} The (apparently) smooth evolution of degree
distribution according to the model with $r= 0.9$.
}
\end{figure}

We performed numerical simulations using these detailed, node-level, ``microscopic" rules (using $r=0.9$) on graphs with
$n=100$ nodes and observed the evolution of several statistical graph properties over time.
In these preliminary numerical experiments, the initial conditions were either empty graphs or
\ER random graphs with a specified value of edge probability, $p$
(of which empty graphs are a special case, corresponding to $p=0$).
It is interesting to consider the evolution of graph properties starting from
an ensemble of initial conditions: \ER graphs with the same edge
probability $p$.
Fig.\ref{fig:dd} shows the evolution of one such property of interest: the
degree distribution of the (nodes of the) graphs, obtained statistically
from a $100$-copy sample.
(The degree of a node in a network is the number of edges connected to the node.)
The figure demonstrates that the degree distribution can be thought of as
a function (ignoring, for the sake of simplicity, the discrete nature of the degree)
that appears to evolve smoothly over time.
Thus, the degree distribution can serve as a potential {\em coarse observable} of the system;
one must, however, carefully investigate whether it is a good candidate for coarse
variable.

\begin{figure}
\includegraphics[width=0.63\textwidth]{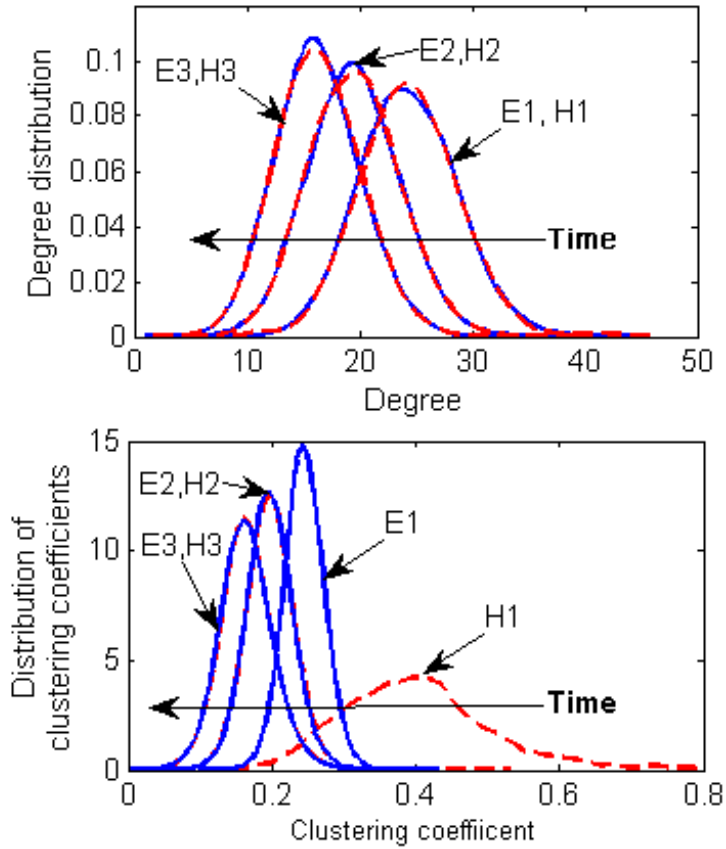}
\caption{\label{fig:sla} Evolution of
(a) degree distribution and (b) triangle distribution, for two
distinct ensembles of graphs.
The solid (blue) curves represent the case where the initial graphs are chosen to be \ER
graphs with $p=0.25$.
The plots corresponding to this case are denoted by the labels, $E1$, $E2$ and $E3$ at times
$0$, $10$ and $20$ respectively.
Note that one time unit in the plot corresponds to $n=100$ iterations of the model.
The dotted (red) curves represent the case where the initial graphs
were created to match the degree distribution of the initial graphs of the previous \ER case
through the Havel-Hakimi algorithm.
The plots corresponding to this second case are denoted by the labels, $H1$, $H2$ and $H3$ at times
$0$, $10$ and $20$ respectively.
}
\end{figure}

In particular, one must test whether it is possible to obtain a
description of the long-term evolution of this observable that is
{\em closed} - that would imply that an accurate reduced model of the system
evolution can, at least in principle, be derived.
%
%
%
%
In other words, if one had measurements of the chosen set of coarse variables (observables)
at particular time,
that information alone should, in principle, be enough to predict
future states of the system (in terms of these variables).
%
%
We reiterate that, depending on the time scales of interest,
different useful reduced models of the same system,
i.e. closures at different levels of coarse description,
may be possible to obtain.

We then proceeded to test whether the degree distribution constitutes a good choice of coarse observable.
For this purpose we constructed initial graphs with identical degree distributions but {\em different higher
order information} (triangle statistics, for instance), evolved the graphs using our model rules
and compared their evolutions in time.
Figures \ref{fig:sla}(a) and \ref{fig:sla}(b) show the evolution of degree distributions and of
clustering coefficient distributions starting from graphs with identical degree distributions, but
distinct distributions of clustering coefficients.
(The clustering coefficient of a node is the ratio of the number of triangles attached to the node
divided by the maximum possible number, given its degree.)
The solid (blue) curves represent the case where the initial graph is an
\ER graph ($p=0.25$).
The dotted (red) curves corresponds to the case where the initial graphs
were created to match the degree distribution of \ER graphs with $p=0.25$;
this was done by sampling $n=100$ numbers (degrees) from the required degree distribution and using the
Havel-Hakimi \cite{Have55remark} algorithm to construct a graph with the sampled sequence of degrees.
The Havel-Hakimi algorithm consists of three iterated steps:
\begin{enumerate}
\item sort the vertices by their degrees ($d_i$) in non-increasing order;
\item select the first vertex ($d_1$) and connect it to the next
$d_1$ vertices; and
\item decrease the value of $d_1$ by $d_1$ (it is now 0) and the value of
the $d_1$ successive degrees by 1.
\end{enumerate}
As an illustration, consider the sorted list of residual degrees of the vertices
(residual degree is the degree of a vertex minus the number of assigned edges
for the same vertex at any given point in the algorithm)
\[ d_1 \geq d_2 \geq ... d_n. \]
Once we connect the first vertex with the next $d_1$ vertices, the residual degree sequence is
\[ (0, d_2-1, d_3-1, ..., d_{d_1+1}-1,d_{d_1+2}, ...). \]
After sorting, the sequence becomes
\[ (d_2-1, d_3-1, ..., d_{d_1+1}-1, ..., 0). \]
These steps are repeated until we get a sequence of zeros, implying that the graph with
the desired degree sequence has been achieved.

For both our test cases, we thus have the same initial degree distribution.
However, since the graphs themselves are different, properties like clustering coefficients
will differ between them, at least initially.
The model was evolved using $100$ copies to obtain smooth distribution functions.
The figures show that the evolution of degree distribution is similar in both cases.
Although the initial clustering coefficient distributions are very different for the two cases, they
{\em quickly} (within a few time steps, compared to the time scale of evolution of the
degree distribution) converge visually to the same function and subsequently coevolve.
This suggests that the clustering coefficient distribution (and possibly other higher order information)
may become quickly {\em slaved} to the evolution of the degree distribution:
triangle statistics (and so also clustering coefficients) evolve at a much faster rate,
and quickly reach a distribution that appears to depend only on the comparatively slowly evolving degree distribution.
These results encourage us to attempt to find a coarse-grained reduction of the system
using a discretization of the degree distribution as the coarse variable(s).

\section{\label{sec:cg}Coarse-graining}

We propose a computer-assisted coarse-graining approach --the Equation-Free (EF) framework \cite{Kevr03equation-free,Kevr04equation-free:}--
to develop and implement a reduced description of the system, using the degree distribution as
the coarse observable.
In this approach, short bursts of simulations at the ``microscopic" (individual node) level are used to
estimate information (such as time-derivatives) pertaining to the coarse variables.
This is accomplished by defining operators that allow us to translate between coarse observables and
consistent detailed, fine realizations.
The transformation from coarse to fine variables is called the {\em lifting} operator ($R$), while the
reverse is called the {\em restriction} operator.
%
%
If we denote the microscopic evolution operator by $\phi_t$, the macroscopic evolution operator can be
defined as
\[ \Phi_t(\cdot) = R \circ \phi_t \circ L(\cdot).\]

As an illustration, we implemented {\em coarse projective integration} \cite{Kevr03equation-free,Kevr04equation-free:}
using the equation-free approach and the degree distribution as the coarse variable.
In coarse projective integration, the system is integrated forward in time using
occasional short bursts of detailed, microscopic simulations at the level of
individual nodes, and the results are
used to  estimate time derivatives at the level of the degree distribution.
In the following discussion, time units are taken to comprise $n=100$ iterations
of the model.
We used \ER random graphs ($p=0.25$) as the initial conditions and
repeatedly performed the following operations:

\begin{enumerate}
\item \textbf{Simulation}: The model is executed initially for $10$ time steps
(in each time step, we perform $n=100$ iterations of the rules of the model).
\item \textbf{Restriction}: The graph degree distributions, $\mu(d)$, are observed from simulations
periodically (at intervals of $2$ time steps).
We stored the degree distribution at discrete values $\mu_i = \mu(d_i)$, where $d_i = 0,1,2,...99$.
The distribution may also be discretized using a smaller number of points and interpolated at intermediate values (we will discuss the possibility of alternative, more parsimonious representations below).
In fact, our plots of degree distributions are smooth interpolations from this $100$ discrete value representation.
\item \textbf{Projection forward in time}: The time-series of the coarse variables over the final segment of the simulation (in our case, the last $3$ observed discretized degree distributions) are used to predict the new values of the coarse variables at a future time (in our case, $10$ time steps
    down the line). This is done on the basis of established numerical integration schemes (in our case,
    simple forward Euler). For the simple model differential equation $dx/dt = f(x)$, the forward Euler scheme would read
           \[  x(t+\Delta t) \approx  x(t) + \Delta t f'(x(t)). \]
    The difference in our case is that the time derivative ($f'(x(t))$ above) does not come form a closed
    formula, but is instead estimated from the recorded recent time series segments.
    In our computations, this time-derivative estimation and projection in time is performed as follows:
    Given the last $3$ observed discretized degree distributions, the associated $3$ {\em cumulative} distribution functions are found, and the degrees, $g_i$, corresponding to uniformly distributed percentile points, $p_i \in \{0,0.01,... 1\}$, $i=1,....,101$,
such that
\[ \int_0^{g_i} \mu(x) dx = p_i. \]
Thus, the pairs of points $(g_i,p_i)$ constitute discrete approximations of the cumulative degree distributions.
The values of these percentile degrees, $g_i$, observed at the last $3$ time steps, are the variables
we actually projected forward in time, estimating the time derivative of the corresponding
forward Euler scheme by
fitting a straight line, and extrapolating it for $10$ further time steps.
When projecting the discretized version of cumulative distributions, one should take care to
retain monotonicity of the predicted (projected)
distributions\cite{morefoot}.
In our simulations we did not encounter such a numerical problem, possibly because
(a) we used several copies to get smoothened degree distributions, and
(b) the projection times were relatively short.

\item \textbf{Lifting}: To restart the simulation, the predicted discretized  distribution
must be transformed into consistent graph realizations.
We accomplish this by using the Havel-Hakimi algorithm; we may have to
sample the projected degree distribution until we draw a graphical sequence of degrees.
(A {\it graphical} sequence is a sequence on non-negative integers that is realizable
as a degree sequence of a graph.)
Checking if a sequence is graphical is performed as a part of the Havel-Hakimi algorithm:
if the algorithm terminates successfully, the sequence is graphical; otherwise, it is not.
In the latter case another sequence is sampled until a graphical one is found.
Once we have these graphs, the procedure repeats: we continue simulations for $10$ more time steps as in stage $1$, keep the last $3$ observations, and so on.
\end{enumerate}

\begin{figure}
\includegraphics[width=0.72\textwidth]{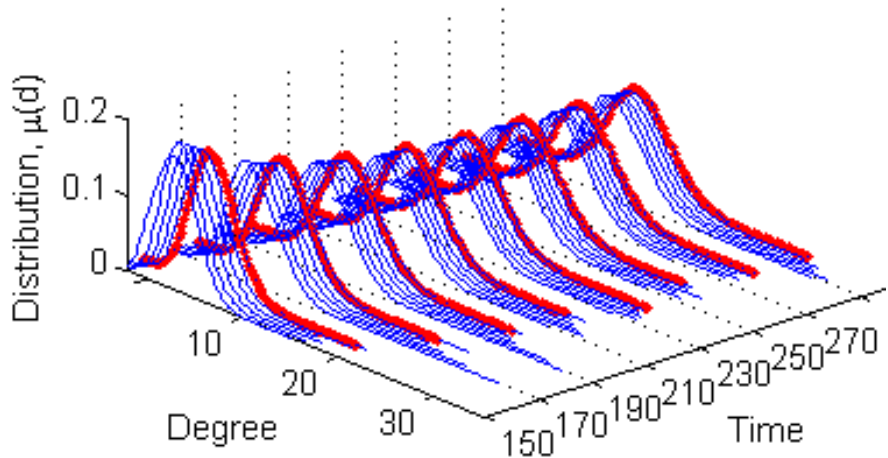}
\caption{\label{fig:cpi} Coarse projective integration (CPI) with
the degree distribution as the coarse variable.
Results from CPI are plotted as (blue) thin lines every $2$ time steps during simulations,
while results from direct simulation are plotted as (red) thick lines every $20$ time steps for comparison.
Note that one time unit in the plot corresponds to $n=100$ iterations of the model.
}
\end{figure}

The results of coarse-projective integration are shown as (blue) thin lines in Fig.~\ref{fig:cpi}
(note that degree distributions are plotted only when we collect simulation data,
and not when we jump in time).
The results from direct integration of the full model are plotted (red, thick lines)
every $20$ time steps for comparison; the evolution of the degree distribution predicted by our coarse projective integration clearly coincides with the full detailed simulation.

\begin{figure}[b]
\includegraphics[width=0.5\textwidth]{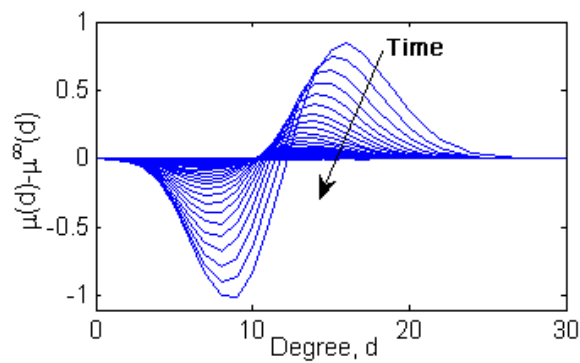}
\caption{\label{fig:app} Decay of the degree distribution close to steady state:
evolution of the difference between the instantaneous degree distribution $\mu(d)$ and
the steady state degree distribution $\mu^{\infty}(d)$.
}
\end{figure}

\begin{figure}[t]
\includegraphics[width=0.6\textwidth]{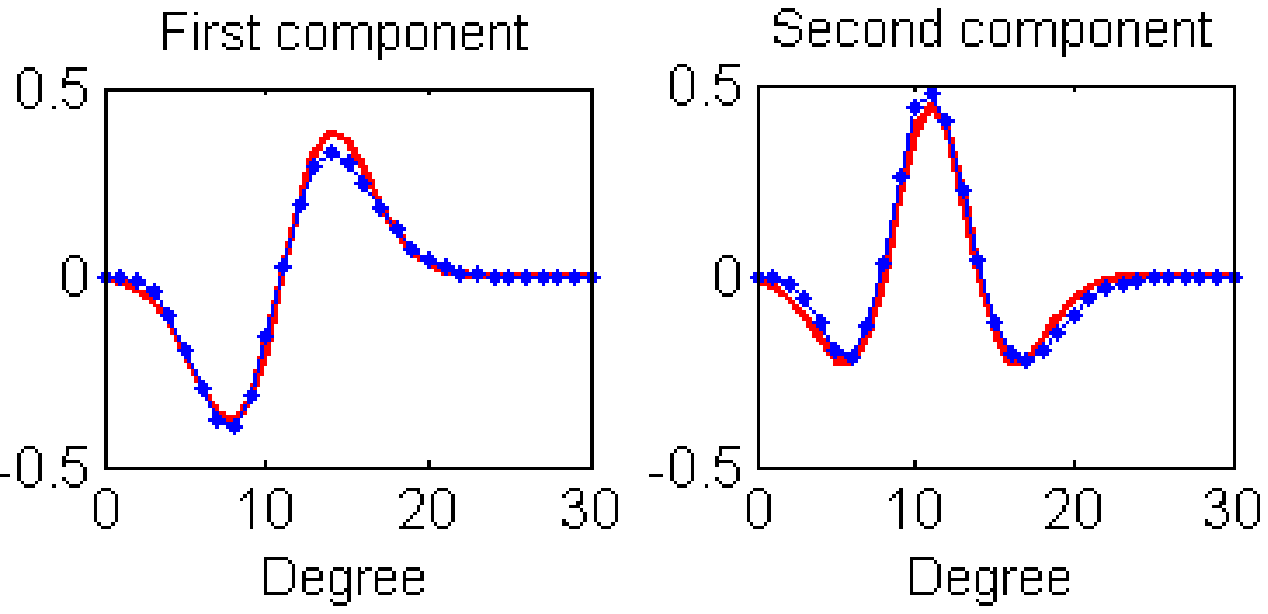}
\caption{\label{fig:appEV} The first two PCA components of
the ensemble of vectors $\mu(d) - \mu^{\infty}(d)$ shown in Fig.~\ref{fig:app}
are plotted as (blue) dotted lines.
These components were found to be well-approximated by the analytical expressions
$e^{(x-11)^2/20}(x-11)/5$ (first component, left)
and $e^{(x-11)^2/20}(x-8)(x-14)/20$ (second component, right),
plotted as (red) solid lines for reference.
}
\end{figure}

We studied the rate of (temporal) convergence of discretized degree distribution as a
given sample network evolution approaches steady (stationary) state.
Fig.~\ref{fig:app} shows the evolution of the difference between the
instantaneous degree distribution $\mu(d)$ close to the steady state and
the steady state degree distribution $\mu^{\infty}(d)$ itself.
We used PCA to evaluate the first few principal components
of the time sequence ($\mu(d) - \mu^{\infty}(d)$).
The first two singular values were found to be $\approx 1.00$ and $0.0175$ respectively.
The vectors corresponding to these first two singular values are plotted as
(blue) dots in Fig.~\ref{fig:appEV}.
These eigenvectors represent directions along which the decay of the degree distribution,
from the given randomly chosen initial condition to the steady state, is the slowest.
We found that the principal components were qualitatively similar for
a wide variety of choices of initial graphs.
For the specific example shown in the figure, the two principal components were found to be well-approximated
by the analytical expressions $e^{(x-11)^2/20}(x-11)/5$ and $e^{(x-11)^2/20}(x-8)(x-14)/20$, which are plotted
as (red) solid lines in the same figure.
The relevance of the qualitative form of these principal components
will be discussed later in Section~\ref{sec:addres}.

In addition to coarse projective integration, we also performed coarse fixed point computations.
Instead of finding the (discretized) stationary degree distribution (coarse fixed point of the evolution)
through direct simulation, one can also obtain it by solving the equation
\[ F(\mu(d)) := \mu(d) - \Phi_{10}(\mu(d)) = 0,\]
where $\Phi_t$ is the coarse time-stepper over $t$ time steps.
We find the roots of $F$ using a damped Newton-Krylov GMRES iteration scheme
\cite{Saad86gmres:,Kell95iterative}.
The standard Newton algorithm updates the value of $\mu(d)$ by
$\mu_{n+1}(d) = \mu_n(d) + \Delta \mu(d)$, where $\Delta \mu(d)$ satisfying
$[DF(\mu_n(d))]\Delta\mu(d) = -F(\mu_n(d)).$
Since $\Phi_{t}(\mu(d))$ is not explicitly available (but can be evaluated
through the coarse time-stepper), its Jacobian cannot be obtained by
analytical differentiation; in the Krylov-based approach the action
of this Jacobian on known vectors (its matrix-vector products) are
estimated through numerical directional derivatives.
Thus, linear (and through them, nonlinear) equations are solved through
iterative {\em matrix-free} computations; these methods are naturally
suitable for equation-free computation, where explicit Jacobians are not
available in closed form.

\begin{figure}
\includegraphics[width=0.45\textwidth]{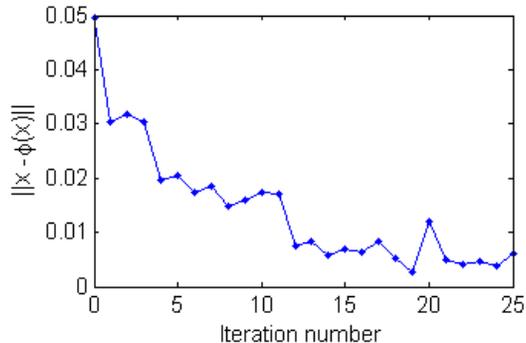}
\caption{\label{fig:nr} Convergence of the Newton-GMRES iterations
in steady state computations: the norm of
the function to be solved for via Newton-GMRES is plotted against the
iteration number.
}
\end{figure}

There are a couple of points worth mentioning about the use of
this general methodology in the case where the unknowns solved for constitute
a (discretized) distribution function  $\mu(d)$: (a) the
$\mu(d)$ vectors that arise should be non-negative and (b) they
should integrate to $1$.
At every iteration of the root-finding algorithm,
these two properties should be preserved.
These conditions on $\mu(d)$ can be stated as conditions on $\Delta \mu(d)$:
$\Delta \mu(d) \leq -\mu_n(d)$ and $\Delta \mu(d)$ should integrate to $0$.
We satisfy the first condition by using a {\it damped} Newton-Krylov method.
The correction term $\Delta \mu(d)$ is scaled
by a constant to ensure non-negativity of the result:
$\mu_{n+1}(d) = \mu_n(d) + c\Delta\mu(d)$ for $c$ sufficiently small.
The second condition is guaranteed by the structure of the {\it Krylov} method itself.
When solving $[DF(\mu_n(d))]\Delta\mu(d) = -F(\mu_n(d))$ using a Krylov method,
the solution $\Delta\mu(d)$ lies in the space spanned by
$\{ F(\mu_n(d)), [DF(\mu_n(d))]\cdot F(\mu_n(d)), ..., [DF(\mu_n(d))]^m\cdot F(\mu_n(d))\}$
for some finite $m<n$.
Because each of the spanning vectors has integral $0$, the update
term $\Delta\mu(d)$ (and hence $c\Delta\mu(d)$, for any constant $c$)
also has integral $0$.
The convergence of this procedure can be shown by plotting $||\mu(d)-\phi(\mu(d))||$
at every iteration as in Fig.~\ref{fig:nr}.

Coarse projective integration and coarse-fixed point algorithms are only two
illustrations of equation-free computation: even
though explicit coarse-grained evolution equations are not available, the
assumption that they {\em in principle} exist helps solve the coarse-grained
model through appropriately designed short bursts of detailed simulation
(also through lifting and restriction).
Many additional computational tasks (e.g. coarse continuation and bifurcation
computations, coarse eigencomputations, coarse controller design and even
optimization) also become possible in this framework \cite{Kevr09equation-free}.
Our computations so far provide numerical corroboration of the possibility of
coarse-graining our network evolution model: a reduced model appears to
close (accurately enough to be usefully predictive) in terms of only the
(discretized) degree distribution.
In what follows, we will provide certain theoretical results to
support our choice of coarse variables and provide some intuition
about the overall coarse-graining approach.

\section{\label{sec:theory}Theoretical justification}

In this Section we discuss theoretical results that
corroborate the observed separation of time scales between the evolution of our
coarse variables (the discretized degree distribution) and
higher order information about the evolving graph statistics (e.g.
the distribution of triangles).
Such information would support our ability to usefully close a
coarse-grained model at the level of degree distributions.

For this simple graph evolution model,
theoretical results for the evolution of edge density and vertex degrees
can be easily derived using basic notions of probability;
this was one of the reasons for choosing this model as our illustration.
In general, for obtaining such results (including results for the evolution
of triangles and more), one makes use of the notion of
dense graph limits, which will be outlined later.

Recall that our model graphs evolve in discrete time steps according to
the rules given in Section \ref{sec:model}.
$G_n(T)$ denotes the graph in $n$ nodes at any discrete iteration, $T = 0,1,...$
$G_n(T,i,j)$ is the entry in the corresponding adjacency matrix
representing the edge between nodes $i$ and $j$.
Let $E(G_n(T))$ represent the set of edges in the graph.

\subsection{\label{ss:edgeden}Evolution of edge density}

We denote by $e_n(T)=\abs{E(G_n(T))}$ the number of edges in the graph at a given
discrete time, $T$.
Let $\bar{\rho}_n(T)$ be the {\em edge density} of the graph, $G_n(T)$ at time $T$:
\begin{equation*}
\bar{\rho}_n(T):= \frac{e_n(T)}{\binom{n}{2}}.
\end{equation*}
Note that the evolution of $e_n(T)$ is itself a Markov
chain, and that it is decoupled from the other variables:
\begin{align}
\label{e_up}
\condprob{e_n(T\!+\!1)-e_n(T)=1}{G_n(T)} & = (1-\bar{\rho}_n(T))\cdot (1-r),
\\
\label{e_stay}
\condprob{e_n(T\!+\!1)-e_n(T)=0}{G_n(T)} & = (1-\bar{\rho}_n(T))\cdot r + \bar{\rho}_n(T)\cdot (1- r),
\\
\label{e_down}
\condprob{e_n(T\!+\!1)-e_n(T)=-1}{G_n(T)} & = \bar{\rho}_n(T))\cdot r.
\end{align}

In order to study the evolution of $\bar{\rho}_n(T)$,
we introduce the continuous time variable $t \in [0,
\infty)$ and let $T=\lfloor \binom{n}{2} t \rfloor $, so that $T+1$
corresponds to $t +\mathrm{d}t$ with
$\mathrm{d}t = \binom{n}{2}^{-1}$.
Let $\rho_n(t):=\bar{\rho}_n(\lfloor \binom{n}{2} t \rfloor))$.
It follows from \eqref{e_up}, \eqref{e_stay}, \eqref{e_down} that we have
\begin{align}
\label{edgedens_condexpect}
\condexpect{ \rho_n(t+\mathrm{d}t) -\rho_n(t)}{ \rho_n(t)} & =(1-r- \rho_n(t)) \mathrm{d}t,
\\
\label{edgedens_condvar}
\condvar{\rho_n(t+\mathrm{d}t) -\rho_n(t)}{ \rho_n(t)} & = \big( (1-\rho_n(t))\rho_n(t) +r \cdot (1-r) \big)  \binom{n}{2}^{-1} \mathrm{d}t.
\end{align}
Letting our process evolve for
 $T \asymp n^2$ steps (i.e.\ $t \asymp 1$), we obtain
\begin{align*}
\expect{ \rho_n(t)-\rho_n(0)} &\stackrel{\eqref{edgedens_condexpect}}{\asymp} T \cdot \mathrm{d}t \asymp 1, \\
\var{\rho_n(t)-\rho_n(0)}   & \stackrel{\eqref{edgedens_condvar}}{\asymp} T \cdot  \binom{n}{2}^{-1} \mathrm{d}t
\asymp n^{-2}.
\end{align*}
Since the variance of $\rho_n(t)-\rho_n(0)$ in the above formulas is much smaller than its expected value if $1 \ll n$,
we can replace $\rho_n(t)$ by
$\expt{\rho_n(t)}$   in \eqref{edgedens_condexpect} without causing significant error.
This leads to the deterministic equation
\begin{equation}\label{ode_edge_density1}
\frac{\mathrm{d}}{\mathrm{d}t} \rho(t) = (1- r)- \rho(t)
\end{equation}
for $\rho(t)$, the limit $\rho_n(t) \to \rho(t)$ as $n \to \infty$,
corresponding to \eqref{edgedens_condexpect}.
Thus
\begin{equation}\label{explicit_edgedens_ode_solution}
\rho(t)=(1-r)+ (\rho(0)-(1-r))e^{-t}
\end{equation}
and $\lim_{t \to \infty} \rho(t) =1-r$.

\begin{figure}
\includegraphics[width=0.45\textwidth]{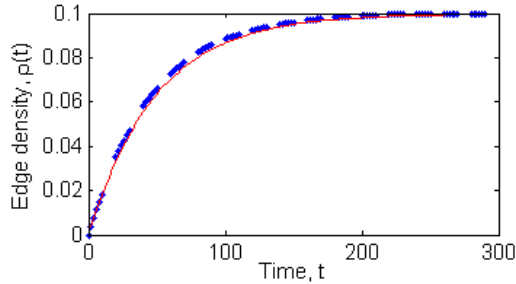}
\caption{\label{fig:cpi1} Coarse projective integration (CPI) of the edge density evolution:
Results obtained from CPI using edge density as the coarse variable are plotted as blue dots
(see text for details).
The trajectory obtained from full direct simulation (red solid line) is plotted for reference.
One time unit in this plot corresponds to $n=100$ iterations of the model.
}
\end{figure}

\smallskip

From \eqref{ode_edge_density1}, it is clear that the evolution of edge
density $\rho(t)$ is {\em truly decoupled from} the evolution of other quantities - one does not need the higher order statistics of the graph to get slaved
to it over a short time period on a sort of {\em slow manifold} for
the reduced model to close.
As mentioned earlier, this represents one type of conditions for which reduced descriptions of the system become possible.
In Section~\ref{sec:cg} we computationally implemented a reduced model
using the entire degree distribution as the coarse observable(s).
\eqref{ode_edge_density1} represents an even more reduced description that meaningfully closes in terms of edge density.
Similar to the degree distribution case, we implement {\em coarse-projective
integration} using now only edge density as the coarse variable.
The single scalar value of the edge density was observed and
recorded during brief bursts of direct simulation;
the time horizons for direct simulation and subsequent projection forward in time were again chosen to be
$10$ time steps each.
The lifting operation in this case involves creating graphs with a given value of edge density.
Our particular implementation of this created graphs where every edge was  selected to be present with a probability equal to the edge density.
The results of coarse projective integration with the empty graph
as initial condition, shown as (blue) dots in Fig.~\ref{fig:cpi1},
agrees visually with the evolution of edge density as observed from direct simulations, shown as a (red) solid line.
This corroborates the closure of a reduced description in terms of edge density only.
We now proceed to derive results supporting our previous observation of closure of a reduced description at the more detailed (less coarse) level of degree distribution.

\subsection{Evolution of the degree of a vertex}

Consider the time evolution of the degree of a node $i \in [n]$,
i.e. the number of other vertices $i$ is connected to.
The order of magnitude of the degree of $i$ is $n$.
Define a normed degree, $D_n(t)$, of a vertex $i$ at scaled time $t$ as
\begin{equation}\label{def_eq_D_n}
  D_n(t):=  \frac{1}{n} \sum_{j=1}^n G_n( \lfloor t \binom{n}{2} \rfloor,i,j);
\end{equation}
we omit the $i$-dependence from the $D_n(t)$ notation for the
sake of simplicity.
Following a derivation similar to the one used in Section \ref{ss:edgeden}, we obtain:

\begin{align}
\label{degree_condexpect}
\condexpect{ D_n(t+\mathrm{d}t) -D_n(t)}{ G_n(t)}
& = \left(1- \left(1+\frac{r}{\rho_n(t)}\right)D_n(t)   \right) \mathrm{d}t,
\\
\label{degree_condvar}
\condvar{D_n(t+\mathrm{d}t) -D_n(t)}{ G_n(t)}
& = \left( 1- D_n(t) + r \frac{D_n(t)}{\rho_n(t)} \right) n^{-1}  \mathrm{d}t.
\end{align}

Now \eqref{degree_condvar} is of smaller order than
\eqref{degree_condexpect} if $n \gg 1$, so that we may replace  $D_n(t)$ by $\expt{D_n(t)}$.
By analogy with the argument of subsection \ref{ss:edgeden} we see that $D_n(t) \to D(t)$ as $n \to \infty$, where $D(t)$ solves the ODE
\begin{equation}\label{ode_degree1}
\frac{\mathrm{d}}{\mathrm{d}t} D(t) = 1- \left(1+\frac{r}{\rho(t)}\right)D(t).
\end{equation}

Substituting the explicit formula \eqref{explicit_edgedens_ode_solution}
into \eqref{ode_degree1} and using the formula for the solution of inhomogenous linear
ODEs, we obtain an explicit solution for $D(t)$:
\begin{equation}\label{degree_explicit}
D(t) = \rho(t) + {e}^{\frac{-t}{1-r}}{\rho(t)}^{\frac{-r}{1-r}}
\left( \rho(0)^{\frac{r}{1-r}} D(0)   - \rho(0)^{\frac{1}{1-r}} \right).
\end{equation}
Clearly, $\lim_{t \to \infty} D(t) = \lim_{t \to \infty} \rho(t) = 1-r$.

Note that the evolution of the degree of a given vertex depends both on its current degree and the current edge density of the graph as a whole.
Yet, if the degree distribution of a graph is given, the edge density of the graph also follows from it.
This clearly supports our observation that the system
closes
at the level described in Section~\ref{sec:cg}:
in terms of the degree distribution.
Approximate differential equations were derived so far to
describe the evolution of (expected values of) the normed degree and of the edge density of graphs.
The next step is to explore the dynamics and influence of higher order information, like triangles.
In order to derive similar results for statistics of triangles,
we take advantage of concepts from the theory of convergent dense graph sequences, of which a rigorous and formal introduction can be found in \cite{Lov06limits}.

\subsection{Convergent dense graph sequences}

Let $G_n(T)$ denote the adjacency matrices of our evolving graphs,
and let the individual entry representing the existence of an edge between nodes $i \in [n]$ and $j \in [n]$ at time $T$ be denoted by $G_n(T,i,j)$.
The notion of dense graph limits will be useful
for describing the time evolution of the statistical properties of $G_n(T)$ when
$n \gg 1$.
%
%
The limit object of a convergent graph sequence is a so-called
\emph{graphon} $W$, where $W: [0,1]^2 \to [0,1]$ is a measurable (but not necessarily continuous) function with the properties
$W(x,y)=W(y,x)$ and $W(x,x)=0$.
Assume that for each $n \in \N$ we have a graph $G_n$ with vertex set $[n]$.
We now informally define the notion of convergence of the sequence $G_n$ to $W$,
i.e.\ $G_n \to W$.

One might heuristically imagine
the adjacency matrix of $G_n$ as a black-and-white television screen
 (a white pixel at position $(i,j)$ represents an edge between vertices $i$ and $j$); a convergent graph
sequence becomes then a sequence of TV sets showing the same picture at higher and higher resolution.
The limiting graphon $W$ will then be the picture seen on the ``perfect TV''
where each point $(x,y) \in [0,1]^2$ is a ``pixel of infinitesimal size''; the local density of black/white
pixels will then give us the impression of shades of grey.
For the precise definition of the so-called cut-distance $\delta(G_n,W)$ between a finite graph and a graphon, see \cite[Section 4]{Lov09very,Lov09very1}.
Note that by \cite[Theorem 6.13]{Lov09very} every graphon $W$ arises as a limit for a convergent graph sequence $G_n$.

Clearly, there exist many adjacency matrices corresponding to different labelings of the nodes of the same graph, and
in the definition of $G_n \to W$
we are allowed to relabel the vertices of $G_n$ (i.e.\ to rearrange the pixels our TV set).
Correspondingly, we are allowed to relabel $[0,1]$ using measure-preserving transformations in order
to obtain equivalent forms of the graphon $W$ (see \cite[Section 3.1]{Lov09very}).
For the purposes of
the present paper, accounting for rearrangements is not required.

If $F$ is the adjacency matrix of a small graph on $k$ nodes,
then we define the homomorphism density $t(F,G_n)$ by
\begin{align}
\label{eqn:homdenG}
t(F,G_n) := \frac{1}{{n\choose k}k!} \!  \sum_{\varphi:[k] \to [n]} \! \ind
\left[ \forall  i, j \in [k] \! : \! F(i,j) \! \leq \! G_n(\varphi(i),
\varphi(j)) \right],
\end{align}
where we sum over all possible injective functions
$\varphi$ from $[k]$ to $[n]$.
$F$ is our {\em test graph} and $t(F,G_n)$ the homomorphism density of $F$ in $G_n$.
%
%

We define the homomorphism density of $F$ in $W$ by
\begin{align}
\label{eqn:homdenW}
t(F,W)=\int_0^1 \! \dots \! \int_0^1 \! \prod_{1 \leq i < j \leq k} \! W(x_i,x_j)^{F(i,j)}  \mathrm{d}x_1 \dots \mathrm{d}x_n.
\end{align}
Denote by $K_k$ the complete graph on $k$ vertices;
for example, $K_2$ is an edge and $K_3$ is a triangle.
Erasing an edge from a triangle gives a ``cherry", a simple graph with three vertices and two edges.

Now, denote by $\rho_n:= \frac{\abs{E(G_n)}}{\binom{n}{2}}$ the edge
density of $G_n$.
It follows from \cite[Section 6.2]{Lov09very} that
$G_n \to W$ implies $t(F,G_n) \to t(F,W)$:
\begin{gather}
\label{edgedens_111}
\rho_W := \lim_{n \to \infty} \rho_n =\lim_{n \to \infty} t(K_2,G_n)
        = t(K_2,W)= \int_0^1 \int_0^1 W(x,y)\mathrm{d}x \mathrm{d}y,
\\
\label{triangledens_111}
 \lim_{n \to \infty} t(K_3,H_n) =t(K_3,W)
             =\int_0^1 \int_0^1 \int_0^1 W(x,y)W(y,z)W(z,x) \mathrm{d}x  \mathrm{d}y  \mathrm{d} z,
\\
\label{cherrydens_111}
 \lim_{n \to \infty} t(\text{cherry},H_n)=t(\text{cherry},W)
 =\int_0^1 \int_0^1 \int_0^1 W(x,y)W(y,z) \mathrm{d}x  \mathrm{d}y  \mathrm{d} z.
\end{gather}
It is the ability to write such equations that makes working with graphons useful for our purposes.
Once the graphon is identified, one can approximate the density of any test graph $F$ in $G_n$, $1 \ll n$
using
expressions similar to Equations \eqref{edgedens_111}, \eqref{triangledens_111}, \eqref{cherrydens_111}.

\subsection{Evolution of the graphon}
If we consider a convergent graph sequence $H_n \to W$, where $H_n$
is a graph on $n$ vertices, and for each $n$ we run our Markov
process $G_n(T)$ with initial state $G_n(0)=H_n$, then
\[ G_n(\lfloor \binom{n}{2} t \rfloor) \to W_t, \]
where $W_t(x,y)$ is the solution of the following ODE:
\begin{equation}\label{ode_graphon}
\frac{\mathrm{d}}{\mathrm{d}t} W_t(x,y)= 1
-\left(1+\frac{r}{\rho(t)}\right)W_t(x,y).
\end{equation}
The heuristic derivation \eqref{heu_graphon_ode} of this formula
is based on
\begin{equation}\label{heu_graphon_prob}
 W_t(x,y) \approx \prob{ G_n( \lfloor \binom{n}{2} t
\rfloor, \lfloor x\cdot n \rfloor , \lfloor y \cdot n \rfloor ) =1}
\end{equation}
  where $T = \lfloor
\binom{n}{2} t \rfloor$, $\mathrm{d}t = \binom{n}{2}^{-1}$, $i
= \lfloor x\cdot n \rfloor$, $j = \lfloor y \cdot n \rfloor$.
Note that
$e(T) \approx \rho(t) \binom{n}{2}$.

\begin{multline} \label{heu_graphon_ode}
 W_{t+\mathrm{d}t}(x,y)-W_t(x,y)=\\
  \condprob{
 G(T+1,i,j)=1}{G(T,i,j)=0}\cdot(1-W_t(x,y))- \\ \condprob{
 G(T+1,i,j)=0}{G(T,i,j)=1} \cdot W_t(x,y)=\\
 \binom{n}{2}^{-1}\cdot(1-W_t(x,y)) - \frac{r}{e(T)} \cdot
 W_t(x,y)=\\
\Big( 1 -\left(1+\frac{r}{\rho(t)}\right)W_t(x,y)\Big)
\mathrm{d}t.
\end{multline}
This results in \eqref{ode_graphon}.

By substituting the explicit formula \eqref{explicit_edgedens_ode_solution}
into \eqref{ode_graphon} and using the integral formula for the solution of
inhomogenous linear ODEs, we obtain an explicit solution for $W_t(x,y)$:
\begin{equation}
W_t(x,y) = \rho(t) + {e}^{\frac{-t}{1-r}}{\rho(t)}^{\frac{-r}{1-r}}
\left( \rho(0)^{\frac{r}{1-r}} W_0(x,y)   - \rho(0)^{\frac{1}{1-r}} \right).
\end{equation}
We have $\lim_{t \to \infty} W_t(x,y) = \lim_{t \to \infty} \rho(t) =1-r$.
Thus for $t \gg 1$ the
graphon becomes almost constant with value $1-r$;
\eqref{heu_graphon_prob} shows that the stationary state of
our graph dynamics looks like an \ER graph with edge density
$1-r$.
In addition, \eqref{ode_graphon} implies that an initially constant graphon will remain constant at future times.
Thus, the family of \ER graphs is {\em an invariant set} of the dynamics of system.

Note that, given the explicit formula for $W_t(x,y)$ we may
obtain an explicit formula for the density of triangles and
cherries in $G_n(T)$ using \eqref{triangledens_111} and
\eqref{cherrydens_111}.
Differential equations for describing the evolution of triangles and
cherry densities at time $t$ can be found directly by differentiating those equations.

\subsection{Evolution of cherry and triangle densities}

Recalling \eqref{cherrydens_111} we have
\[ t(\text{cherry}, W_t)=\int_0^1 \int_0^1 \int_0^1
W_t(x,y)W_t(y,z) \mathrm{d}x  \mathrm{d}y  \mathrm{d} z. \]
Differentiating \eqref{cherrydens_111} with respect to time and using
the graphon evolution result of \eqref{ode_graphon}, we get
\begin{multline}\label{cherry_density_ode}
\frac{\mathrm{d}}{\mathrm{d}t} t(\text{cherry},W_t)=
2 \!  \int_0^1
\! \int_0^1 \! \int_0^1 \! \left( 1- \left( \! 1+\frac{r}{\rho(t) }
\! \right) \! \! \! W_t(x,y) \right) W_t(y,z) \mathrm{d}x  \mathrm{d}y
\mathrm{d} z =\\
2 \rho(t) -2\left( 1+\frac{r}{\rho(t) }
\right)t(\text{cherry},W_t).
\end{multline}

\smallskip
\noindent
An ODE describing the time evolution of the density of triangles
(in terms of the edge density and the cherry density) can be
similarly derived:
\begin{equation}\label{triangle_density_ode}
\frac{\mathrm{d}}{\mathrm{d}t} t(K_3,W_t)=
3~t(\text{cherry},W_t) -3\left( \! 1+ \! \frac{r}{\rho(t)}
\! \right)t(K_3,W_t).
\end{equation}

\subsection{\label{ss:conrate}Convergence rates}

Now that we have equations describing the statistics of degrees,
cherries and triangles, we can find out the rates at which these quantities
converge to their steady (stationary) states.
A function $a(t)$ converges to another function $b(t)$ at rate $\alpha$ if
\[ \lim_{t \to \infty} \frac{1}{t}\log( \abs{a(t)-b(t)})=-\alpha.
\]
In order to establish this, it is enough to prove that
\[ \lim_{t \to \infty} \frac{1}{\abs{a(t)-b(t)}} \frac{\mathrm{d}}{\mathrm{d}t}\abs{a(t)-b(t)} = -\alpha.
\]

\smallskip
\noindent

From \eqref{explicit_edgedens_ode_solution} we can directly see that the edge densities of our graphs
converge to the steady state value of $1-r$  at a rate $\alpha = 1$.
This can also be shown by using \eqref{ode_edge_density1} to write the following equation:
\[ \frac{\mathrm{d}}{\mathrm{d}t} \left( \rho(t)-(1-r) \right)\;  = \; - \left( \rho(t)-(1-r)
\right). \]

We now show that the normed degree $D(t)$ of  a vertex converges
to $\rho(t)$ at a faster rate.
From \eqref{ode_edge_density1},\eqref{ode_degree1} we obtain

\begin{equation}\label{degree_convR}
\frac{\mathrm{d}}{\mathrm{d}t} \left( D(t)-\rho(t) \right)=
-\left( 1+\frac{r}{\rho(t) } \right)\left( D(t)-\rho(t) \right).
\end{equation}
Thus $\alpha=\lim_{t \to \infty} \left( 1+\frac{r}{\rho(t) }
\right)=\frac{1}{1-r}$ in this case. Note that $\alpha=\frac{1}{1-r}$
also follows from the explicit formula \eqref{degree_explicit}.
For example, if $r=0.9$ then $\alpha=10.$

\medskip
\noindent
From \eqref{ode_graphon}, we similarly obtain that for any $x,y \in
[0,1]$ the function $W_t(x,y)$ converges to $\rho(t)$ at a rate
$\alpha=\frac{1}{1-r}$.

\medskip

If the graphon $W_t$ evolves according to \eqref{ode_graphon} and
$\rho(t) = \int_0^1 \int_0^1 W_t(x,y)  \mathrm{d}x
\mathrm{d}y$ then $\rho(t)$ solves \eqref{ode_edge_density1}.
If we let $\hat{W}_t(x,y)=\rho(t)$ for any $x,y \in [0,1]$ then
$\hat{W}_t$ also solves \eqref{ode_graphon}.
Thus, the set of constant graphons is invariant under the dynamics.
We now show evidence that this ``invariant manifold" is actually attracting:
\begin{multline}\label{cherry_convR}
\frac{\mathrm{d}}{\mathrm{d}t} \left( t(\text{cherry}, W_t) -
t(\text{cherry}, \hat{W}_t) \right)
\stackrel{\eqref{cherrydens_111}}{=}
\frac{\mathrm{d}}{\mathrm{d}t} \left( t(\text{cherry}, W_t) -
\rho(t)^2 \right)\stackrel{\eqref{cherry_density_ode},
\eqref{ode_edge_density1}}{=} \\
2 \rho(t) -2\left( 1+\frac{r}{\rho(t) }
\right)t(\text{cherry},W_t) -2 \rho(t) (1-r-\rho(t))=\\
-2\left( 1+\frac{r}{\rho(t) } \right)\left( t(\text{cherry}, W_t)
- t(\text{cherry}, \hat{W}_t) \right).
 \end{multline}
This implies that $t(\text{cherry}, W_t)$ converges to $t(\text{cherry},
\hat{W}_t)$ at rate $\alpha=2\frac{1}{1-r}$.
If $r=0.9$ then $\alpha=20$ for this case.
Thus, if we evolve graphs that already possess the steady state values of their edge density, their
cherries will converge to their steady state value twice as fast as the rate at which normed degrees evolve to their corresponding steady state.

We now consider triangle density evolution.
Let $W^1$ and $W^2$ be two distinct graphons with the same values of edge and
cherry densities:
\[ t(K_2,W^1\!)\!=\!t(K_2,W^2\!) \text{ and } t(\text{cherry},W^1\!)\!=\!t(\text{cherry},W^2\!) \]
If we let $W^1_t$ and $W^2_t$ evolve according to the dynamics \eqref{ode_graphon}
with initial states $W^1$ and $W^2$, respectively, then
by \eqref{ode_edge_density1} and \eqref{cherry_density_ode} we have
\begin{equation} \label{eq_same_dens_cherry}
t(K_2,W_t^1\!)\!=t(K_2,W_t^2\!), \; t(\text{cherry},W_t^1\!)\!=\!t(\text{cherry},W_t^2\!)
 \end{equation}
for all $t \geq 0$.
The fact that the density of edges and cherries coincide for
$W^1_t$ and $W^2_t$ ``helps" the densities of triangles  in $W^1_t$
and $W^2_t$ to converge to each other even more rapidly:
\begin{equation}\label{triangle_convR}
\frac{\mathrm{d}}{\mathrm{d}t}
\left( t(K_3, W^1_t) - t(K_3, W^2_t)
\right)\stackrel{\eqref{triangle_density_ode}}{=}
-3\left(1+\frac{r}{\rho(t) } \right)\left( t(K_3, W^1_t) - t(K_3, W^2_t) \right).
\end{equation}
Thus the rate of convergence of the relative triangle density is $\alpha=3\frac{1}{1-r}$: for $r=0.9$ this works out to be $30$.
This result, in particular, explains why we observed an
apparent slaving of the triangles, as discussed in
in Fig. \ref{fig:sla}.
The (blue) solid and (red) dotted curves there showed the evolution of
two graphs with the same degree distribution.
Graphs with the same degree sequence also have the same number of
edges and cherries, which implies \eqref{eq_same_dens_cherry}.
The number of triangles in these two graphs (corresponding to the two cases
in Fig.\ref{fig:sla}) converge to each other three times as fast as the
rate at which the degree distributions ultimately evolve.

\begin{figure}[t]
\includegraphics[width=0.9\textwidth]{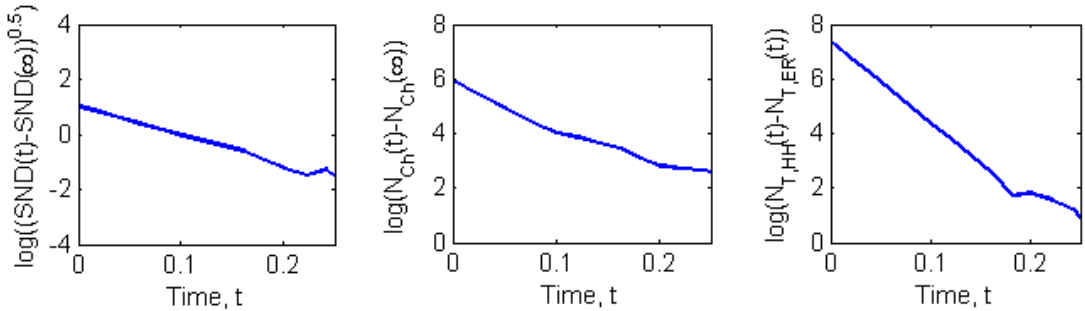}
\caption{\label{fig:conv} Evolution of the logarithms of quantities related to
(a) degree, (b) cherries and (c) triangles.
The initial conditions for the three cases are explained in the text.
SND denotes the sum of squares of the normed sequence of degrees of all the $100$ nodes.
$N_{Ch}$ denotes the number of cherries in the graph.
$N_{T,ER}$ and $N_{T,HH}$ denote the number of triangles in two graph evolutions starting with an
\ER random graph with $p=0.1$ and a graph created using the Havel-Hakimi algorithm
respectively, both with the same degree distribution.
Note that one time unit in this plot corresponds to $\binom{n}{2}=4950$ iterations of the model.
Slopes of the three lines are $-10.44$, $-18.49$ and $-29.98$ respectively, which
are in close agreement with theoretical results of $-10$, $-20$ and $-30$ respectively.
}
\end{figure}

We now estimate convergence rates from direct simulation results,
confirming the validity of our approach and approximations {\em even} for relatively small systems:
in these results  $n=100$ nodes and we simulate the model using a value of $0.9$ for the parameter $r$.
Figure \ref{fig:conv} shows how information about degrees, cherries and triangles converge to their
corresponding steady (stationary) states.
Note that, in all these cases, the time $t$ is scaled so that one time unit comprises $\binom{n}{2}$ iterations of the model.
Logarithms of quantities (defined in the caption of the figure)
related to these statistics are plotted in the y-axis versus time.
For producing the first two plots (corresponding to degrees and cherries), the initial
graphs were created by first sampling from a normal degree distribution with mean $10$
and standard deviation of $1$.
For comparison, the steady state graphs have a degree distribution whose mean and
standard deviation were evaluated to be $\approx 10$ and $\approx 3$ respectively.
Thus, the initial graphs have the same mean degree (and hence the same edge density)
as the steady state graphs, but differ from these steady state graphs in their actual detailed degree distribution.
Since the graphs are initialized with the steady state edge density, this edge density remains
close to its steady state value at all times.
From \eqref{degree_convR} and \eqref{cherry_convR}, the convergence rates for the
quantities in the first two figures are expected to be $10$ and $20$ respectively.
This successfully matches the numerical convergence rates (the absolute values of the slopes of evolutions)
of the terms related to degrees and cherries in the figure: they
are seen to be $10.44$ and $18.49$ respectively.
For the third plot, containing information about triangles, we simulate the model {\em from two
different initial graphs} with the same degree distribution {\em but different number of triangles}.
The first simulation is initialized with an \ER random graph with $p=0.1$.
For the second case, we created initial graphs using the Havel-Hakimi algorithm, using the degree sequence of the first case as input.
Since graphs with the same degree sequence also have the same number of
edges and cherries, \eqref{eq_same_dens_cherry} is satisfied.
Hence from \eqref{triangle_convR}, we expect a convergence rate of $30$ for the
{\em relative} number of triangles {\em between these two graphs}, which successfully matches the
numerically computed value of $29.98$.
Thus, although the theoretical results are in principle accurate only at the limit of
very large graphs, all the numerical values we computed using graphs with only $100$ nodes are remarkably close to the limiting theoretical values.

\section{\label{sec:addres}Some additional theoretical results}

\subsection{An SDE for the degree of a vertex}

In the previous section we approximated the evolution of the normed degree through a {\em deterministic} ODE, arguing for the relative smallness of the order of magnitude of the variance of what is really a stochastic process.
In order to now describe the evolution of the stochastic process $D_n(t)$ at a finer level, we approximate it by a stochastic differential equation (SDE) rather than an ODE:
\begin{equation}\label{eq_sde_for_degree}
\mathrm{d} D_n(t) = \left(1- D_n(t)-r\frac{D_n(t)}{\rho_n(t)} \right)  \mathrm{d}t +
\sqrt{ 1- D_n(t)+r\frac{D_n(t)}{\rho_n(t)} } \cdot
n^{-1/2} \mathrm{d} W_t,
\end{equation}
where $W_t$ now denotes the standard Brownian motion.
To rationalize the choice of such an approximation, we observe that
the solution of \eqref{eq_sde_for_degree} satisfies
 \eqref{degree_condexpect} and \eqref{degree_condvar} and that
the $D_n(t)$ defined by \eqref{def_eq_D_n} satisfies
$|D_n(t+\mathrm{d}t)-D_n(t)| \leq \frac{1}{n}$, i.e.\ the trajectory of
$D_n(t)$ is very close to being continuous.
One can then suggest that it is appropriate (and even natural) to use the Brownian motion in \eqref{eq_sde_for_degree} as a driving function.
For rigorous results validating the SDE approximation of discrete stochastic processes,
see \cite{Ispa10note}.

We are interested in the fluctuations of $D_n(t)$ around its
expected value $\tilde{D}_n(t)=\expect{D(t)}$.
Using \eqref{degree_condexpect} we can see that $\tilde{D}_n(t)$ approximately solves
the ODE
\begin{equation}\label{eq_ode_for_expected_degree}
 \mathrm{d} \tilde{D}_n(t) = \left(1- \tilde{D}_n(t)-r\frac{\tilde{D}_n(t)}{\rho_n(t)} \right)  \mathrm{d}t.
\end{equation}
If we define $X(t):= n^{1/2}(D(t)-\tilde{D}_n(t))$ then by
\eqref{eq_sde_for_degree} and \eqref{eq_ode_for_expected_degree}  $X(t)$ solves the SDE
\begin{equation} \label{SDE_for_degree}
\mathrm{d} X(t)=- \left(1+ \frac{r}{\rho_n(t)} \right) X(t) \mathrm{d}t + \sqrt{ 1- D_n(t)+
r\frac{D_n(t)}{\rho_n(t)} }  \mathrm{d} W_t.
\end{equation}
If $n$ is big enough, then ${\rho}_n(t) \approx \rho(t)$
and $D_n(t) \approx \tilde{D}_n(t) \approx D(t)$,
so we may use the deterministic $\rho(t)$ and $D(t)$
in the right-hand side SDE of $X(t)$ without causing much error.

Since $\rho(t)$ and $D(t)$ are known and explicit
(c.f.\ \eqref{explicit_edgedens_ode_solution}, \eqref{degree_explicit}),
\eqref{SDE_for_degree} is a linear SDE, i.e. an Ornstein-Uhlenbeck
process with time-dependent drift and diffusion coefficient.
From this, it follows that $X(t)$ can be approximated by a Gaussian process with mean $0$
and an explicit formula for the variance at time $t$.
If we let $t \to \infty$ then $\rho(t) \to 1-r$ and
$D(t) \to 1-r$, so \eqref{SDE_for_degree} becomes
\begin{equation}\label{sde_degree_X_OU}
 \mathrm{d} X(t)=- \frac{1}{1-r} X(t)
\mathrm{d}t + \sqrt{ 2 r  }
\mathrm{d} W_t,
\end{equation}
an OU process.
The variance of the stationary distribution of this Markov process
can be expressed using the drift and diffusion coefficients and
it is normally distributed:
\begin{align}\label{stationary_degree_normal_X}
X(\infty) \sim \mathcal{N}\left( m=0, \sigma^2 = \frac{ 2 r }{2
\cdot \frac{1}{1-r}}\right)
= \mathcal{N}\left( m=0, \sigma =
\sqrt{(1-r)r} \right)
\end{align}
Now $D_n(t) = \bar{D}_n(t)+ n^{-1/2} X(t)$, from which we get:
\begin{equation}\label{stationary_degree_normal}
D_n(\infty) \sim \mathcal{N} \Big(
 m= 1-r, \; \sigma = \frac{1}{\sqrt{n}} \sqrt{r(1-r)} \Big)
 \end{equation}

It is worth noting that similar results can be derived for the edge density by defining
$Y(t):= \binom{n}{2}^{1/2}(\rho(t)-\tilde{\rho}(t))$.
$Y(t)$ solves the SDE
\begin{equation}\label{SDE_for_edgedensity}
\mathrm{d} Y(t)=-Y(t)
\mathrm{d}t + \sqrt{ (1-\rho(t))\rho(t) +r \cdot (1-r) }
\mathrm{d} W_t
\end{equation}
This is analogous to \eqref{SDE_for_degree}.
If we let $t \to \infty$ then $\bar{\rho}(t) \to 1-r$,
so \eqref{SDE_for_edgedensity} becomes the following OU process:
\begin{equation*}
\mathrm{d} Y(t)=-Y(t)
\mathrm{d}t + \sqrt{ 2 r \cdot (1-r) }
\mathrm{d} W_t.
\end{equation*}

\subsection{A PDE for the evolution of the normed degree probability distribution}

If we consider the SDE \eqref{sde_degree_X_OU} and denote the
probability density function of $X(t)$ by $P(t,x)$, then $P(t,x)$
solves the Fokker-Planck (or Kolmogorov-forward) equation:
\begin{equation}\label{pde}
\frac{\partial}
{\partial t} P(t,x)= \frac{1}{1-r} \frac{\partial}{\partial x} \big( x \cdot  P(t,x) \big) + r \frac{\partial^2}{\partial x^2} P(t,x).
\end{equation}
A simple argument is that, when $1 \ll n$,
the trajectories of the evolving degrees of vertices $i$ and $j$
show little correlation,
since the source of randomness for the
degree evolution for distinct vertices is almost disjoint: they have at most one edge in common.
It then follows that observations of the time evolution of
the empirical degree distribution histograms (for the entire graph)
may be well approximated by solutions of the Fokker-Planck equation \eqref{pde} (for
the degree probability density of a single node).

Solving the eigenvalue-eigenfunction problem
corresponding to the differential operator on the right-hand side
of \eqref{pde} gives rise to the Hermite differential
equation, whose eigenfunctions are the Hermite functions.
In particular, the first three eigenvalues and eigenfunctions are
\[
\begin{array}{rl}
\lambda_0=0, & \quad  f_0(x) = \exp\left( -\frac{1}{2} \frac{x^2}{r\cdot(1-r)}  \right) \\
\lambda_1=-\frac{1}{1-r}, & \quad  f_1(x) = \exp\left( -\frac{1}{2} \frac{x^2}{r\cdot(1-r)}  \right)\cdot x \\
\lambda_2=-\frac{2}{1-r}, & \quad
f_2(x) = \exp\left( -\frac{1}{2} \frac{x^2}{r\cdot(1-r)}  \right)\cdot( r - \frac{x^2}{1-r} )\\
\end{array}
\]

\begin{itemize}
\item $f_0(x)$  is (a
    constant times) the density function of \eqref{stationary_degree_normal_X}
    which is also the stationary solution of \eqref{pde}.
\item $f_1(x)$ represents
    the direction along which the decay of a generic initial distribution $P(0,x)$
    to $P(\infty,x)$ is the slowest:
    \begin{equation*}
    \lim_{t \to \infty} \exp(\frac{t}{1-r}) \cdot \left(
    P(t,x)-P(\infty,x) \right) =
    c \cdot f_1(x)
    \end{equation*}
\item $f_2(x)$ represents
    the direction along which the decay of $P(0,x)$
    to $P(\infty,x)$ is the slowest, if $P(0,x)$ is a generic {\em even} function of $x$.
\end{itemize}

These formulas corroborate the plots in Fig.~\ref{fig:appEV}, even though the
assumptions made in order to derive the theoretical results are valid only at the limit of very
large graphs.
It is interesting to note that the leading principal components of the decay of the empirical degree distribution histograms to steady state, which we found
to be well approximated by the functions $e^{(x-11)^2/20}(x-11)/5$ and $e^{(x-11)^2/20}(x-8)(x-14)/20$,
can also be very well matched to the Fokker-Planck eigenfunctions $f_1(x)$ and $f_2(x)$ by shifting and rescaling coordinates.

\section{\label{sec:conc}Summary and Conclusions}

In this paper, we have demonstrated a computational framework for coarse-graining evolutionary problems involving networks.
We illustrated our methodology using a specific model with simple, random evolution rules.
The proposed methodology applies, in principle, to
{\em any} network evolutionary model with an inherent separation of time scales between the
evolutions of a few important coarse variables, and the remaining slaved variables (observables).
For our illustrative model, we were able to analytically derive certain
theoretical results, justifying our choice of coarse variables and quantifying the observed time scale
separation.
We used the notion of dense graph limits to formulate some of our arguments for successful computational coarse-graining.
It is conceivable that some of the theoretical tools used here might be
useful in deriving insights in other dynamic network models.
We emphasize, however, that for the right problems our coarse-graining procedure will work irrespective of
whether one is able to analytically derive such supporting theoretical results.

The generality of the approach raises other important general
questions in the area of complex networks.
We mentioned earlier that a critical step is the
identification of suitable coarse variables.
There are at least two open questions that need to be addressed in that regard:
\begin{enumerate}
  \item How does one find the appropriate coarse variables for a given model?
  \item Once suitable coarse variables are identified,
  how does one solve the problem of constructing networks with
  prescribed values of the chosen variables?
\end{enumerate}

The second question is more concrete, and hence, we will address it first.
For certain specific properties, there exist well-known algorithms to construct graphs
with specified values of those properties.
For instance, for constructing networks with a given degree distribution,
we repeatedly used the Havel-Hakimi algorithm \cite{Have55remark,Haki62realizability}.
Alternative approaches to construct graphs with a specified degree distribution include, among others,
the configuration model \cite{Boll80probabilistic,Worm80some}, the Chung-Lu model \cite{Chun02connected},
and a sequential importance sampling algorithm \cite{Blit06sequential}.
Beyond the degree distribution, there are only a few properties for which
standard approaches have been established for constructing graphs with specified values of those properties.
For example, algorithms that create graphs with the following properties can be found in the literature:
given degree-degree distribution \cite{Doro02how},
given degree distribution {\em and} average clustering coefficient \cite{Kim04performance}, and
given degree distribution {\em and} degree dependent clustering \cite{Serr05tuning}.
For other properties (or combinations of properties),
more generalized mathematical programming approaches (e.g. \cite{Goun11generation})
could potentially be used to solve the network generation problem.

The first question, however, requires more new ideas.
For the example presented here, coarse-graining was originally based
on computational model observations, and preceded the derivation of theoretical results.
In general, the motivation for good coarse variables can come from standard heuristics,
intuition about the model under consideration or observations of evolution
of statistical quantities of the model.
However, smart use of data mining tools such as diffusion maps \cite{Nadl06diffusion,Lafo06diffusion}
might provide answers in a more generic sense.
This would require the definition of a useful metric quantifying the distance between
nearby graphs (see e.g. \cite{Borg06counting,Vis08graph}).
Automatic data mining to extract good coarse variables from model observations
is, in some sense, a holy grail of model reduction methods.

\begin{acknowledgments}

This work was partially supported by DTRA (HDTRA1-07-1-0005)
and by the US DOE (DE-SC0002097). Parts of this work
are contained in the Ph.D. Thesis of K.A.B.; it is a pleasure for I.G.K. to acknowledge
several discussions with Professor L. Lovasz, and to thank him for bringing us together with B.R.
\end{acknowledgments}


\begin{thebibliography}{99}

\bibitem{Albe02statistical}
     \newblock R. Albert and A. L. Barab\'{a}si,
     \newblock \emph{Statistical mechanics of complex networks},
     \newblock Rev. Mod. Phys., \textbf{74} (2002), 47--97.

\bibitem{Aren06synchronizationa}
     \newblock A. Arenas, A. D\'{i}az-Guilera and C. J. P\'{e}rez-Vicente,
     \newblock \emph{Synchronization reveals topological scales in complex networks},
     \newblock Phys. Rev. Lett., \textbf{96} (2006), 114102.

\bibitem{Barr08dynamical}
     \newblock A. Barrat, M. Barthelemy and A. Vespignani.
     \newblock ``Dynamical processes on complex networks,"
     \newblock Cambridge University Press, 2008.

\bibitem{Binz09topology}
     \newblock T.~Binzegger, R.~J. Douglas and K.~A.~C. Martin,
     \newblock \emph{Topology and dynamics of the canonical circuit of cat v1},
     \newblock Neural Networks, \textbf{22} (2009), 1071--1078.

\bibitem{Blit06sequential}
\newblock J. Blitzstein and P. Diaconis,
\newblock ``A sequential importance sampling algorithm for generating random
  graphs with prescribed degrees,"
\newblock Technical report, 2006. Available from: \url{http://www.people.fas.harvard.edu/~blitz/BlitzsteinDiaconisGraphAlgorithm.pdf}.

\bibitem{Bocc06complex}
     \newblock S.~Boccaletti, V.~Latora, Y.~Moreno, M.~Chavez and D.-U. Hwang,
     \newblock \emph{Complex networks: Structure and dynamics},
     \newblock Physics Reports, \textbf{424} (2006), 175--308.

\bibitem{Bold07equation-free}
     \newblock K.~A. Bold, Y.~Zou, I.~G. Kevrekidis and M.~A. Hensonevrekidis,
     \newblock \emph{An equation-free approach to analyzing heterogeneous cell population dynamics},
     \newblock J. Math. Biol., \textbf{55} (2007), 331--352.

\bibitem{Boll80probabilistic}
     \newblock B.~Bollobas,
     \newblock \emph{A probabilistic proof of an asymptotic formula for the number of labelled regular graphs},
     \newblock European J. Combin., \textbf{1} (1980), 311--316.

\bibitem{Borg06counting}
     \newblock C. Borgs, J. Chayes, L. Lov\'{a}sz, V. S\'{o}s and K. Vesztergombi,
     \newblock ``Topics in Discrete Mathematics: Algorithms and Combinatorics,"
     \newblock Springer, Berlin, 2006.

\bibitem{Chen04from}
     \newblock L.~Chen, P.~G. Debenedetti, C.~W. Gear and I.~G. Kevrekidis,
     \newblock \emph{From molecular dynamics to coarse self-similar solutions: a simple
  example using equation-free computation},
     \newblock J. Non-Newton Fluid, \textbf{120} (2004), 215--223,

\bibitem{Chun02connected}
     \newblock F. Chung and L. Lu,
     \newblock \emph{Connected components in random graphs with given expected degree sequences},
     \newblock Ann. Comb., \textbf{6} (2002), 125--145.

\bibitem{Doro02how}
     \newblock S.~N. Dorogovtsev, J.~F.~F. Mendes and A.~N. Samukhin,
     \newblock \emph{How to construct a correlated net},
     \newblock  eprint, arXiv:cond-mat/0206131.

\bibitem{Falo99power-law}
     \newblock M. Faloutsos, P. Faloutsos and C. Faloutsos,
     \newblock \emph{On power-law relationships of the internet topology},
     \newblock In SIGCOMM, (1999), 251--262.

\bibitem{Goun11generation}
     \newblock C. Gounaris, K. Rajendran, I. Kevrekidis and C. Floudas,
     \newblock \emph{Generation of networks with prescribed degree-dependent clustering},
     \newblock Optim. Lett., \textbf{5} (2011), 435--451.

\bibitem{Haki62realizability}
     \newblock S.~L. Hakimi,
     \newblock \emph{On realizability of a set of integers as degrees of the vertices of a
  linear graph. I},
     \newblock J. Soc. Ind. Appl. Math., \textbf{10} (1962), 496--506.

\bibitem{Have55remark}
     \newblock V.~Havel,
     \newblock \emph{A remark on the existence of finite graphs. (czech)},
     \newblock Casopis Pest. Mat., \textbf{80} (1955), 477--480.

\bibitem{Ispa10note}
     \newblock M. {Isp{\'a}ny} and G. {Pap},
     \newblock \emph{A note on weak convergence of random step processes},
     \newblock Acta Mathematica Hungarica, \textbf{126} (2010), 381--395.

\bibitem{Kell95iterative}
     \newblock C.~T. Kelley,
     \newblock \emph{Iterative Methods for Linear and Nonlinear Equations},
     \newblock number 16 in Frontiers in Applied Mathematics, SIAM, Philadelphia (1995).

\bibitem{Kevr04equation-free:}
     \newblock I.~G. Kevrekidis, C.~W. Gear and G. Hummer,
     \newblock \emph{Equation-free: The computer-aided analysis of complex multiscale systems},
     \newblock AIChE Journal, \textbf{50} (2004), 1346--1355.

\bibitem{Kevr03equation-free}
     \newblock I.~G. Kevrekidis, C.~W. Gear, J.~M. Hyman, P.~G. Kevrekidis, O. Runborg and C. Theodoropoulos,
     \newblock \emph{Equation-free, coarse-grained multiscale computation: enabling
  microscopic simulators to perform system-level analysis},
     \newblock Commun. Math. Sci., \textbf{1} (2003), 715--762.

\bibitem{Kevr09equation-free}
     \newblock I.~G. Kevrekidis and G. Samaey,
     \newblock \emph{Equation-free multiscale computation: algorithms and applications},
     \newblock Annu. Rev. Phys. Chem., \textbf{60} (2009), 321--344.

\bibitem{Kim04performance}
     \newblock B.~J. Kim,
     \newblock \emph{Performance of networks of artificial neurons: the role of
  clustering},
     \newblock Phys. Rev. E, \textbf{69} (2004), 045101.

\bibitem{Lafo06diffusion}
     \newblock S. Lafon and A. B. Lee,
     \newblock \emph{Diffusion maps and coarse-graining: A unified framework for dimensionality reduction, graph partitioning and data set parameterization},
     \newblock IEEE T. Pattern Anal., \textbf{28} (2006), 1393--1403.

\bibitem{Lain09dynamics}
     \newblock C.~R. Laing,
     \newblock \emph{The dynamics of chimera states in heterogeneous kuramoto networks},
     \newblock Physica D, \textbf{238} (2009), 1569--1588.

\bibitem{Lov09very}
     \newblock L.~{Lov\'{a}sz},
     \newblock \emph{Very large graphs},
     \newblock  eprint, arXiv:0902.0132.

\bibitem{Lov09very1}
     \newblock L.~{Lov\'{a}sz},
     \newblock \emph{Very large graphs, in: Current Developments in Mathematics 2008},
     \newblock International Press, Somerville, MA, 2009.

\bibitem{Lov06limits}
     \newblock L. Lov\'{a}sz and B. Szegedy,
     \newblock \emph{Limits of dense graph sequences},
     \newblock J. Comb. Theory Ser. B, \textbf{96} (2006), 933--957.

\bibitem{Moon07heterogeneous}
     \newblock S. J. Moon, B.~Nabet, N.~E. Leonard, S.~A. Levin and I.~G.~Kevrekidis,
     \newblock \emph{Heterogeneous animal group models and their group-level alignment
  dynamics: An equation-free approach},
     \newblock J. Theor. Biol., \textbf{246} (2007), 100--112.

\bibitem{Nadl06diffusion}
     \newblock B. Nadler, S. Lafon, R. R. Coifman and I. G. Kevrekidis,
     \newblock \emph{Diffusion maps, spectral clustering and reaction coordinates of dynamical systems},
     \newblock Appl. Comput. Harmon. A., \textbf{21} (2006), 113--127.

\bibitem{Newm03structure}
     \newblock M.~E.~J. Newman,
     \newblock \emph{The structure and function of complex networks},
     \newblock SIAM Review, \textbf{45} (2003), 167--256.

\bibitem{Newm02random}
     \newblock M.~E~J Newman, D.~J. Watts and S.~H. Strogatz,
     \newblock \emph{Random graph models of social networks},
     \newblock Proc. Natl. Acad. Sci., \textbf{1} (2002), 2566--2572.

\bibitem{Newm06structure}
    \newblock M.~E.~J. Newman, A.~L. Barab\'{a}si, and D.~J. Watts,
    \newblock \emph{The Structure and Dynamics of Networks},
    \newblock Princeton University Press, 2006.

\bibitem{Raje11coarse}
     \newblock K. Rajendran and I.~G. Kevrekidis,
     \newblock \emph{Coarse graining the dynamics of heterogeneous oscillators in networks
  with spectral gaps},
     \newblock Phys. Rev. E, \textbf{84} (2011), 036708.

\bibitem{Saad86gmres:}
     \newblock Y. Saad and M.~H. Schultz,
     \newblock \emph{GMRES: A generalized minimal residual algorithm for solving
  nonsymmetric linear systems},
     \newblock SIAM J. Sci. Stat. Comp., \textbf{7} (1986), 856--869.

\bibitem{Serr05tuning}
     \newblock M. A. Serrano and M. Bogun\'{a},
     \newblock \emph{Tuning clustering in random networks with arbitrary degree
  distributions},
     \newblock Phys. Rev. E, \textbf{72} (2005), 036133.

\bibitem{Toiv09comparative}
     \newblock R. Toivonen, L. Kovanen, M. Kivel\"{a}, J. Onnela, J. Saram\"{a}ki and K. Kaski,
     \newblock \emph{A comparative study of social network models: Network evolution
  models and nodal attribute models},
     \newblock Soc. Networks, \textbf{31} (2009), 240--254.

\bibitem{Vis08graph}
     \newblock S.~V.~N. Vishwanathan, K.~M. Borgwardt, I. Risi Kondor and N.~N. {Schraudolph},
     \newblock \emph{Graph Kernels},
     \newblock  eprint, arXiv:0807.0093.

\bibitem{Worm80some}
     \newblock N.~C. Wormald,
     \newblock \emph{Some problems in the enumeration of labelled graphs},
     \newblock B. Aust. Math. Soc., \textbf{21} (1980), 159--160.

\bibitem{morefoot}
     Here, we discuss an alternative method of projection in time:
     %
     For each of the three simulated discrete cumulative distributions (defined by $\bar{g}=[g_1,g_2,...]$),
     the median degree $g_m$ (i.e., $g_i$ which corresponds to $p_i = 0.5$) is evaluated
     and the shifted cumulative distribution $\bar{g}=[g_1-g_m,g_2-g_m,...]$ is computed.
     %
     Thus, we have the new vector $\bar{g}$ at each of the three time steps.
     %
     PCA (Principal Component Analysis) is then used to find the first two normalized principal components, $v_1$ and $v_2$, of this  ensemble of
     three vectors.
     %
     The coefficients of the projection of our three $\bar{g}$ vectors along these first two principal components are then computed - we denote them
     by $\lambda_1$ and $\lambda_2$ respectively.
     %
     The median degree ($g_m$) and these two coefficients ($\lambda_1$ and $\lambda_2$)
     are then (independently) projected forward (in a coarse projective forward Euler
     step) through the linear extrapolation described above.
     %
     From these three projected values, the predicted discretized degree distribution $\mu(d)$
     at the projected time ($10$ time steps down the line) can be constructed.
\end{thebibliography}
\end{document}